\documentclass[useAMS,usenatbib,usegraphicx]{mn2e}

\title[Evolutionary constraints on the planet host $\epsilon$\,Reticulum]
{Evolutionary constraints on the planet-hosting subgiant $\epsilon$\,Reticulum from its white dwarf companion}

\author[J. Farihi et al.]{J. Farihi$^1$, M. R. Burleigh$^1$, J. B. Holberg$^2$, S. L. Casewell$^1$, and M. A. Barstow$^1$\\
$^1$Department of Physics \& Astronomy, University of Leicester, Leicester LE1 7RH, UK;
	jf123@star.le.ac.uk\\
$^2$Lunar and Planetary Laboratory, University of Arizona, Tucson AZ 85721-0063, USA}

\begin{document}

\date{}

\maketitle

\label{firstpage}

\begin{abstract}

The planet hosting and Sirius-type binary system $\epsilon$\,Reticulum is examined from the 
perspective of its more evolved white dwarf secondary.  The stellar parameters are determined 
from a combination of Balmer line spectroscopy, gravitational redshift, and solid angle.  These
three methods conspire to yield the most accurate physical description of the companion to date:
$T_{\rm eff}=15\,310\pm350$\,K and $M=0.60\pm0.02\,M_{\odot}$.  Post-main sequence mass 
loss indicates the current binary separation has increased by a factor of 1.6 from its primordial 
state when the current primary was forming its planet(s), implying $a_0\geq150$ AU and 
constraining stable planets to within $15-20$\,AU for a binary eccentricity of $e=0.5$.  Almost 
80 years have passed since the first detection of the stellar companion, and marginal orbital 
motion may be apparent in the binary, suggesting a near edge-on configuration with $i\ga70\degr$, 
albeit with substantial uncertainty.  If correct, and all known bodies are coplanar, the mass of the 
planet HD\,27442b is bound between 1.6 and 1.7 $M_{\rm J}$.

A search for photospheric metals in the DA white dwarf yields no detections, and hence there is 
no clear signature of an extant planetary system orbiting the previously more massive secondary.  
However, if the white dwarf mass derived via spectral fitting is correct, its evolution could have 
been influenced by interactions with inner planets during the asymptotic giant branch.  Based on 
the frequency of giant planets and circumstellar debris as a function of stellar mass, it is unlikely 
that the primordial primary would be void of planets, given at least one orbiting its less massive 
sibling.

\end{abstract}


\section{INTRODUCTION}

Given that most stars form in binaries or multiples, understanding planet formation and 
evolution in the presence of two or more stars is fundamental.  Furthermore, planets within 
binary star systems provide empirical constraints for formation theories and the dynamical
models necessary to yield long-term, stable orbits \citep{hol99,hep74}.
	
$\epsilon$\,Reticulum, or HD\,27442A is a K2\,IV star with a gas giant planet in 
a 428 day orbit whose minimum mass is 1.6\,$M_{\rm J}$ \citep{but06,but01}, and a 
spatially resolved, faint stellar companion first detected nearly 80 years ago and found to 
be in the same relative position three and a half decades later \citep{mas01}.  The common 
proper motion of HD\,27442B was established by \citet{rag06}, recognizing it as a stellar 
companion to a planet host star.  The nature of the secondary star was first constrained by 
\citet{cha06}, finding that only a white dwarf was consistent with the optical and near-infrared 
photometry at the 18.2\,pc trigonometric parallax distance to the primary.  The hydrogen-rich, 
degenerate nature of HD\,27442B was confirmed via optical spectroscopy that revealed the 
distinct, pressure-broadened H$\alpha$ absorption profile typical of DA white dwarfs 
\citep{mug07}.  Hence this system is a nearby Sirius-type binary hosting at least one planet.

The white dwarf is not yet listed in the catalog of \citet{mcc08,mcc99}, nor in the 20\,pc 
sample of \citet{hol08}, but should be designated WD\,0415$-$594 based on its B1950
coordinates, as is convention.  However, it is discussed in \citet{hol09} and \citet{sio09},
both of which consider the fraction of white dwarfs present in binary systems.  HD\,27442
is among the eleven known Sirius-type systems within 20 pc and is one of three within this 
distance known to also harbor planets; the others being GJ\,86 and HD\,147513 \citep{des07}.

Combined with the compact stellar radius implied by the distance to the primary, the spectrum 
obtained by \citet{mug07} clearly establishes the secondary as a white dwarf.  However, the 
shape and strength of H$\alpha$ is relatively degenerate over a broad range of $T_{\rm eff}$ 
and $\log g$, while spectroscopy of the higher Balmer lines can effectively separate and 
uniquely determine these two parameters. \citep{ber92}.  This paper presents a determination of 
$T_{\rm eff}$ and $\log g$ for HD\,27442B via high-resolution optical-ultraviolet spectroscopy of 
the entire detected Balmer series up to H$_9$.  This new and accurate information is harnessed,
and combined with reliable system parameters, to place limits on the stellar masses and binary 
separation during the epoch of planet formation, and to trace out the likely post-main sequence 
dynamical history.

\section{OBSERVATIONS \& DATA}

\subsection{UVES Echelle Spectroscopy}

HD\,27442B was observed on 2008 October 20 at Cerro Paranal with the 8.2\,m Very Large 
Telescope using the Ultraviolet and Visual Echelle Spectrograph (UVES; \citealt{dek00}) 
on Unit Telescope 2.  Echelle spectroscopy was performed over the two detectors covering 
wavelengths 3200\,\AA \ to 6650\,\AA\ using a standard dichroic configuration with central 
wavelengths $\lambda_c=3900/5640$\,\AA, resulting in two narrow gaps in spectral coverage 
near 4550 and 5650\,\AA.  A single exposure of 1800\,s was obtained using a slit width of 
$0\farcs5$ with $1\times1$ binning, resulting in a nominal resolving power of $R\approx80
\,000$ in both the UV-Blue and Red arms of the instrument.  The featureless white dwarf 
WD\,0000$-$345 (also LHS\,1008) was observed for program 382.D-0804(A) as a spectral 
standard on 2008 October 8 using an identical UVES setup but with $2\times2$ binning.

\begin{figure}
\includegraphics[width=124mm]{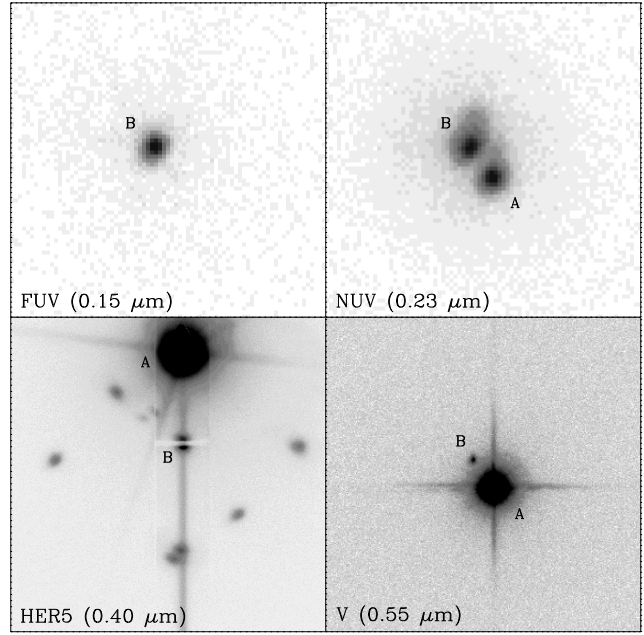}
\caption{{\em Upper panels:  GALEX} far- and near-ultraviolet images of HD\,27442AB.  In 
the epoch 2007.77, near-ultraviolet frame the pair appear separated by $14\farcs0$ at position 
angle $35\fdg1$.  {\em Lower right:}  CTIO 0.9\,m $V$-band CCD image of HD\,27442AB in
a 1\,s exposure.  These three frames are all are North up, East left and approximately $2'$ on 
a side.  {\em Lower left:}  UVES acquisition frame of the science target in the slit, taken through
a below-slit filter with central wavelength near 0.40\,$\mu$m.  The acquisition image is roughly
$45''$ square, and oriented along the binary axis, at approximately $235\degr$.  The CCD 
bleed from the saturated primary does not correspond to stray light entering the slit.
\label{fig1}}
\end{figure}

Figure \ref{fig1} reveals that no difficulty was encountered obtaining the secondary spectrum 
in the presence of its much brighter primary; neither the extended stellar image halo nor the 
telescope support diffraction spikes fall across the slit.  Therefore the spectrum of HD\,27442B
is uncontaminated by the light of the bright primary.  However, in addition to the fact that science
target and standard star were observed on separate nights, there was also a significant difference 
in airmass between these observations, $\Delta \sec{z}\approx0.45$.  Hence the overall shape of 
the reduced spectrum is potentially skewed by differential extinction, owing to both night-to-night 
and airmass variations between science and standard targets, especially at the shortest 
wavelengths.

%%%FIGURE1%%%

\begin{figure*}
\includegraphics[width=172mm]{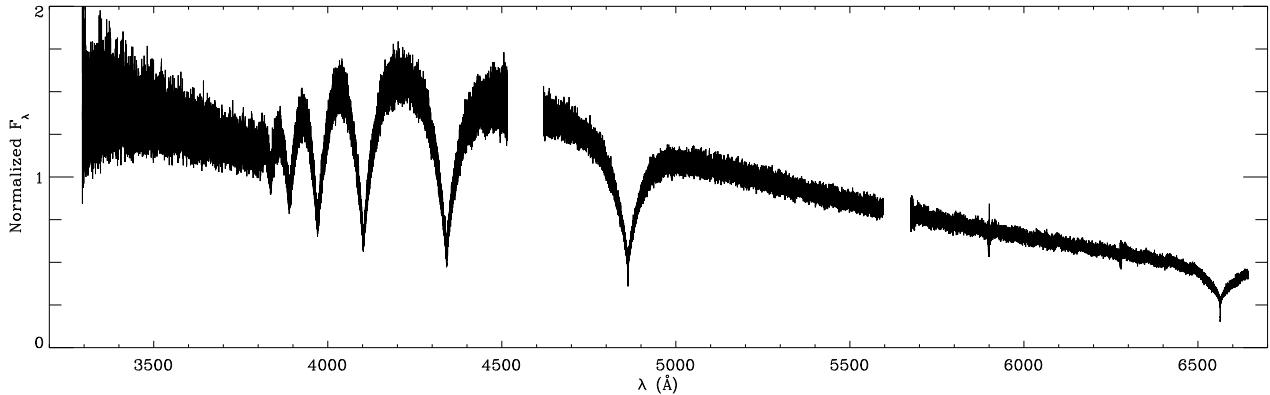}
\caption{UVES spectrum of the white dwarf HD\,27442B.  The science target data were 
interpolated and rebinned in the spectral direction to match the wavelength solution of the DC 
white dwarf observed for sensitivity and flux calibration purposes, but the data are otherwise 
unsmoothed.  The features near 5900\AA \ and 6300\AA \ are detector artifact and telluric 
absorption residuals, respectively.  The quasi-periodic pattern in the spectrum is due to a
light path difference between the flat-field lamp and the sky; this interference is not fully 
correctable, especially at high S/N.
\label{fig2}}
\end{figure*}

The echelle data were processed with the UVES pipeline version 4.3.0 using {\sc gasgano}, 
including cosmic ray masking, flat fielding, wavelength calibration, order merging, and distilled 
using optimal aperture extraction.  The science target data were then interpolated in wavelength 
to match the solution for the spectral standard, divided by the standard spectrum, and multiplied 
by an appropriate temperature blackbody for relative flux calibration.  It was found at this stage 
that the three spectral orders did not match at their adjacent ends, and corrective factors were 
applied to these residual offsets prior re-normalization.  The fully reduced spectrum still contains 
ripples that are often associated with UVES data, but which could not be removed in this high 
signal-to-noise (S/N) dataset; the quality control parameter {\sf rplpar}, which should be much 
less than 5 according to the UVES manual, was less than 0.6 for all extracted orders.  The 
normalized spectrum of HD\,27442B is displayed in Figure \ref{fig2}.

%%%FIGURE2%%%

\subsection{Supplementary Data}

Perhaps unsurprisingly, there is a spatially well-resolved detection of the white dwarf 
companion in {\em Galaxy Evolution Explorer} ({\em GALEX}; \citealt{mar05}) images of
HD\,27442, shown in  Figure \ref{fig1}.  In fact, the white dwarf outshines the K2 subgiant star 
in the near-ultraviolet at 2300\,\AA, and is the only emission source seen in the far-ultraviolet 
at 1500\,\AA.  Roughly speaking, the binary offset apparent in the {\em GALEX} image is
similar to those determined by \citet{cha06} and \citet{mug07}, as well those given in the 
Washington Double Star Catalog \citep{mas01} based on observations made 45 and 80 
years prior.

As of this writing, the SIMBAD IRCS coordinates for HD\,27442B are essentially no different 
than that of the primary.  Taking a nominal offset from this position and extrapolating back to 
epoch 2000.0 using the well-measured proper motion of the system, gives a J2000 position 
for the white dwarf of $04^{\rm h} 16^{\rm m} 30\fs0 -59\degr 17' 58"$ (and a B1950 position 
of $04^{\rm h} 15^{\rm m} 37\fs7 -59\degr 25' 17''$, hence WD\,0415$-$594).

There is a corresponding faint {\rm ROSAT} source centered $18''$ from the J2000 position 
of the primary (1RXS\,J041631.2$-$591815; \citealt{vog00}).  The X-ray detection has an $18''$ 
position error, and hence could be associated with either binary component.  Furthermore, the 
hardness ratio indicates the source is only emitting in the $0.1-0.4$ keV band, but only white 
dwarfs hotter than 25\,000\,K are significant sources of soft X-rays \citep{odw03}.  Observations 
taken with the {\em International Ultraviolet Explorer} ({\em IUE}: \citealt{bog78}) of HD\,27442A 
reveal emission features over $1000-2000$\,\AA \ region, notably very strong Lyman $\alpha$.  
Therefore, it is likely that coronal emission from the cool primary is the source of the detected 
X-rays (i.e., flaring).

Table \ref{tbl1} lists available ultraviolet, optical, and near-infrared photometry for the two
stellar binary components.  The only available near-infrared $JK$ photometry of HD\,27442B
has substantial uncertainty \citet{cha06}, as it is based on the highly saturated 2MASS photometry 
for HD\,27442A.  However, the secondary $H$-band photometry of \citet{mug07} was derived using 
non-adaptive optics images and standard photometric calibration with multiple 2MASS sources in 
the image field (M. Mugrauer 2009, private communication), and should therefore be reliable.

\section{ANALYSIS}

\subsection{White Dwarf Parameter Determination}

\subsubsection{Balmer Spectroscopy}

The spectrum of the white dwarf was fitted using a grid of (LTE) pure hydrogen atmosphere models 
as described in \citet{koe10}.  Comparison between the models and the UVES data was performed
using the spectral fitting routine {\sc fitsb2}, as described in some detail in \citet{cas09}.  Results of 
the spectral analysis yield $T_{\rm eff}=15310\pm20$\,K and $\log\,g=7.88\pm0.01$, including only 
the formal errors of the Balmer line fitting procedure.

There is a significant body of literature on the effective temperatures and surface gravities of 
hydrogen-rich white dwarfs as derived via Balmer line spectroscopy \citep{ber95,ber92}.  Recently, 
two studies have obtained multiple spectra for each of hundreds of DA stars, permitting an assessment 
of errors beyond those derived from the spectral fitting procedure (which are merely a function of S/N):
one using single-order, low-resolution data \citep{lie05}, and the other with high-resolution, echelle 
spectroscopy \citep{koe09}.  Typical standard deviations in these studies are found to be 1.2 and
2.3\% in $T_{\rm eff}$, 0.04 and 0.08\,dex in $\log\,g$, with the higher variance observed in echelle
data, its intrinsic complexity the likely culprit \citep{koe09}.

Particularly appropriate are the results of the SPY survey \citep{nap03}, which observed hundreds 
of DA white dwarfs with UVES using an instrumental configuration and calibration procedure identical 
to that adopted for HD\,27442B.  \citet{koe09} report a systematic offset of $-0.08$\,dex in $\log\,g$ and 
$+1.2$\% in $T_{\rm eff}$ for 85 single, and well-behaved DA white dwarfs in common with \citep{lie05}.  
Despite the high S/N spectrum of HD\,27442B, the aforementioned results imply realistic parameters of 
$T_{\rm eff}=15310\pm350$\,K, $\log\,g=7.88\pm0.08$, with the possibility that the surface gravity and
mass have been systematically under-determined.

%%%TABLE1%%%
\begin{table}
\begin{center}
\caption{Observed and Derived Properties of HD\,27442\label{tbl1}} 
\begin{tabular}{@{}lrr@{}}
\hline
\hline
Component								&B				&A\\
\hline
Photometry:&&\\

FUV (AB\,mag)								&12.7			&...\\
NUV (AB\,mag)								&12.6			&13.0\\
$U$ (mag)								&...				&6.6\\
$V$ (mag)								&12.5			&4.4\\
$I$ (mag)									&...				&3.4\\
$H$ (mag)								&12.9			&2.1\\
\hline
Astrometry:&&\\
$d_{\pi}$ (pc)								&...				&18.2\\
$v_{\rm rad}$ (km\,s$^{-1}$) 					&...				&$+28.7$\\
$(\mu_{\alpha},\mu_{\delta})$ (mas\,yr$^{-1}$)		&...				&$(-48.0,-167.8)$\\
($U,V,W$)	 (km\,s$^{-1}$)						&...				&($-25,-17,-12$)\\
\hline
Parameters:&&\\
SpT										&DA3.3			&K2\,IV\\
$T_{\rm eff}$ (K)							&$15\,310\pm350$	&4850\\
$\log\,g\,({\rm cm\,s}^{-2})$					&$7.98\pm0.02$	&3.78\\
$M$($M_{\odot}$)							&$0.60\pm0.02$	&1.54\\
$M_{\rm ms}$($M_{\odot}$)					&1.9				&...\\
Cooling Age (Gyr)							&0.2				&...\\
MS Age (Gyr)								&1.3				&...\\
Total Age (Gyr)								&1.5				&2.8\\

\hline
\end{tabular}
\end{center}

{\em Note}.  The white dwarf mass and cooling age are based on the models of \citet{fon01}, 
with a main-sequence lifetime estimated using the formulae of \citet{hur02}.  Remaining table 
entries are based on various catalogs and literature sources \citep{mug07,tak07,but06,mar05,
mas01,per97,bes90}.  The 2MASS data for the bright primary is heavily saturated; \citet{gez99} 
gives $J=2.57$\,mag and $K=1.97$\,mag, consistent with the adopted $H$-band magnitude.

\end{table}

\subsubsection{Gravitational Redshift}

The apparent velocity of the secondary was determined using both Gaussian and Lorentzian fits 
to the relatively sharp, non-LTE cores of H$\alpha$ and H$\beta$, yielding $+58.0\pm0.5$\,km\,s$
^{-1}$, including an instrumental stability uncertainty of 0.4\,km\,s$^{-1}$ (R. Napiwotzki 2011, 
private communication).  This observed velocity is the sum of four components:

\begin{equation}
v_{\rm app} = \gamma + v_{\rm orb} + v_{\rm grav} + v_{\rm bary}
\label{eqn0}
\end{equation}

\noindent The systemic velocity ($\gamma$) is subsumed by the total radial velocity ($\gamma 
+ v_{\rm orb}$) of the primary star, for which there are at least four similar measurements published 
between 1913 and 1928, with $+29.3\pm0.5$\,km\,s$^{-1}$ listed in the General Catalog of Stellar 
Radial Velocities \citep{wil53}.  Much more recently, one of the precision radial velocity monitoring 
campaigns has measured $+28.7\pm1.6$\,km\,s$^{-1}$ (the quoted error is over-conservative as the 
instrumental stability is better than 0.3\,km\,s$^{-1}$; J. Jenkins 2010, private communication).  The 
barycentric velocity towards the science target on the UVES observation date was $-1.0$\,km\,s$^
{-1}$.

From the current projected separation of the pair, the maximum orbital speed is 2.8\,km\,s$^{-1}$,
while averaging over $\sin{i}$ for the unknown orbital inclination predicts the observed velocity 
should be within $\pm1.8$\,km\,s$^{-1}$.  This latter value is adopted as the $2\sigma_{\rm orb}$ 
error, with $3\sigma_{\rm orb}=2.7$\,km\,s$^{-1}$ corresponding to the worst case scenario.  Thus, 
adding in quadrature the $1\sigma$ uncertainties in the measured apparent velocity (0.5\,km\,s$
^{-1}$), measured radial velocity (1.6\,km\,s$^{-1}$), and orbital velocity (0.9\,km\,s$^{-1}$) of the 
system yields $v_{\rm grav}=+30.3\pm1.9$\,km\,s$^{-1}$.  This value suggests a  higher mass for 
the white dwarf than predicted by the model fit to the Balmer lines.  

\subsubsection{Solid Angle}

\begin{figure}
\includegraphics[width=86mm]{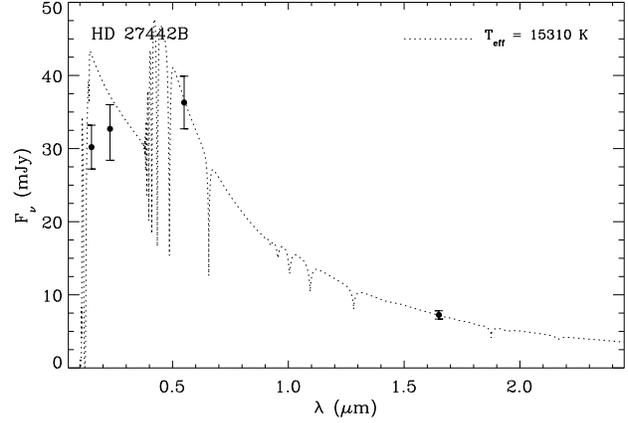}
\caption{Spectral energy distribution of HD\,27442B.  The model effective temperature broadly 
matches the optical through near-infrared color, though the {\em GALEX} ultraviolet photometry 
is somewhat discrepant.
\label{fig3}}
\end{figure}

A third, largely independent check of the white dwarf radius can be made because the system 
has a precise parallax measurement of $\pi=54.83\pm0.15$\,mas \citep{van07}.  The solid angle 
subtended by the star in the sky, and thus $(R/D)^2$, can be determined by fitting a flux-calibrated 
spectral model to the observed stellar flux at the Earth.  Figure \ref{fig3} displays the best fit of the 
spectroscopically-derived model to the data, weighted heavily towards the observed $H=12.871
\pm0.085$\,mag \citep{mug07}, as it is the only published (non-adaptive optics) photometric datum 
with a reliable error.  From the fitting process, a range of acceptable fits are found for 
$R=0.01320\pm0.00045$\,$R_{\odot}$.

%%%FIGURE3%%%

\subsubsection{Adopted Values and Evolutionary Considerations}
 
\begin{figure}
\includegraphics[width=86mm]{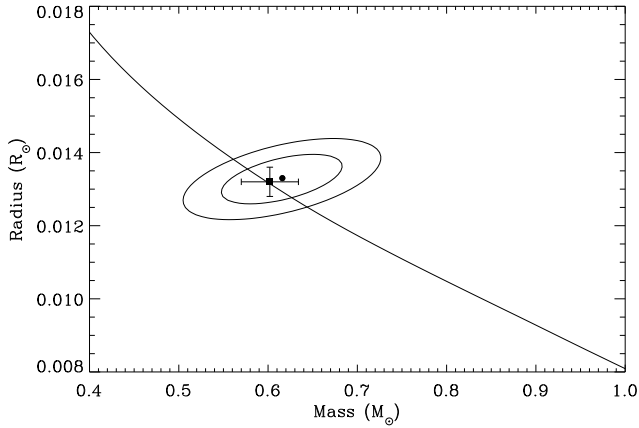}
\caption{Theoretical mass-radius relation for a 15\,310\,K DA white dwarf (diagonal curve) 
plotted with the semi-empirical constraints derived for the photometric radius, surface gravity, 
and gravitational redshift measurements discussed in \S3.1.  The resulting minimum $\chi^2
=1.65$ is indicated by a filled circle along with 1 and $2\sigma$ contours.  For comparison, 
the weighted average mass and radius is plotted as a square with $2\sigma$ 
error bars.
\label{fig4}}
\end{figure}

For many white dwarfs where $T_{\rm eff}$ and $\log\,g$ are determined spectroscopically, 
a mass is routinely calculated based on an assumed, theoretical mass-radius relation.  In 
the case of HD\,27442B, however, there are two additional observations that can be used to 
determine masses and radii: the photometric solid angle (and parallax), and the gravitational 
redshift.  Because the solid angle constrains only $R$, the surface gravity only $M/R^2$, and 
the gravitational redshift only $M/R$, it is necessary to use a mass-radius relation to determine 
both parameters.  Each separate constraint can be used with a mass-radius relation to make 
independent estimates of mass and radius and associated uncertainties.  These independent 
results need not agree precisely but ought to be reasonably consistent.  Table \ref{tbl2} lists the 
results for the three observational constraints on the stellar mass and radius.  Figure \ref{fig4} plots 
the mass-radius relation\footnote{http://www.astro.umontreal.ca/$\sim$bergeron/CoolingModels}
\citep{hol06,fon01} adopted here. 

The mass-radius of HD\,27442B derived via gravitational redshift and solid angle agree 
rather well, predicting a mass roughly 10\% higher than the spectroscopic method.  A higher 
white dwarf mass is superior when viewed in the context of the binary history.  A current mass 
of 0.60\,$M_{\odot}$ corresponds to a progenitor mass near 1.9\,$M_{\odot}$ \citep{kal08}, while 
the spectroscopically-derived value of 0.55\,$M_{\odot}$ implies a progenitor that was (likely, but 
not certainly) less massive than HD\,27442A (1.6\,$M_{\odot}$; \citealt{tak07}).  Given the close
agreement between the stellar parameters derived via redshift and solid angle, plus the possible 
paradox implied by the lower mass derived via spectroscopy, it is tempting to favor higher mass 
values.  Furthermore, the finding of \citep{koe09} that Balmer line fitting in UVES spectra may 
underestimate white dwarf surface gravity makes the lower mass suspect.

If the mass derived via the Balmer line fitting is correct, some process has reduced the white dwarf 
mass to a value 10\% lower than expected via single star evolution.  In this case, a main-sequence
star of $M>1.6\,M_{\odot}$ would have descended into a 0.55\,$M_{\odot}$ remnant.  Since the binary 
has always been sufficiently wide as to preclude mass transfer, this potential conundrum cannot be 
resolved via an Algol-type history.  One possibility is enhanced mass loss along the first ascent giant
branch, during the growth of the degenerate core mass, due to the interactions with one or more inner, 
giant planets \citep{sie99a,sie99b}.  While speculative, it is consistent with the presence of a planetary
system within the binary, and reasonable expectations for planet formation at intermediate-mass stars 
\citep{bow10,lov07,jon07}.

Another, model-independent way of considering the various possible solutions to the radius and 
mass associated with observed photometry, surface gravity and gravitational redshift of HD\,27442B 
is to treat all of these as independent constraints on the stellar radius as a function of mass.  Under 
this type of analysis the radius and mass are determined without explicit reference to any mass-radius 
relation and the most likely mass and radius with associated uncertainties are determined from a 
minimum $\chi^2$ calculation.  In Figure \ref{fig4} is shown how these joint constraints relate to the 
appropriate mass-radius-relation.   From this method the minimum $\chi^2$ corresponds to a mass 
and radius of 0.616\,$M_{\odot}$ and 0.0133\,$R_{\odot}$.  While this result is consistent with the 
expected mass-radius relationship and the weighted average of the results in Table \ref{tbl2}, it yields 
a somewhat higher mass compared to the weighted average of 0.602\,$M_{\odot}$.  This is primarily 
driven by the fact that the mass-radius based determinations will lie closer to the mass-radius relation 
from which they are derived.  In summary, both the mass-radius based determinations and the minimum 
$\chi^2$ calculation give consistent results, where the $\chi^2$ $1\sigma$ contour encloses the 
mass-radius relation for masses between 0.577 and 0.628\,$M_{\odot}$.

%%%TABLE2%%%
\begin{table}
\begin{center}
\caption{Mass-Radius Determinations for HD\,27442B\label{tbl2}} 
\begin{tabular}{@{}lcc@{}}
\hline
\hline

Method						&Radius					&Mass\\
							&($R_{\odot}$)				&($M_{\odot}$)\\
 
\hline

Balmer Spectroscopy			&$0.0141\pm0.0006$		&$0.547\pm0.043$\\
Gravitational Redshift			&$0.0129\pm0.0003$		&$0.616\pm0.022$\\
Solid Angle					&$0.0132\pm0.0005$		&$0.599\pm0.027$\\

\hline

Weighted Average				&$0.0132\pm0.0002$		&$0.602\pm0.016$\\
$\chi^2$ Minimization			&$0.0133\pm0.0007$		&$0.616\pm0.068$\\

\hline
\end{tabular}
\end{center}

{\em Note}.  For $M_B\la0.58$\,$M_{\odot}$, the predicted main-sequence progenitor mass
is lower than the mass of the less evolved primary.

\end{table}

\subsection{Evidence of Two Planetary Systems?}

Because the main-sequence progenitor of the extant white dwarf secondary was more massive 
than the current planet-bearing primary, it is logical that a substantial protoplanetary disk would 
have orbited HD\,27442B.  Hence it is possible that planets would have formed first, and perhaps 
more readily, at HD\,27442B, corroborated by studies that show giant planets and signatures of 
planetary systems are found more often at higher mass stars \citep{bow10,tri08,su06}.  Currently, 
any putative planets orbiting beyond 50\,AU at either stellar component are unstable, yet each 
star may retain planets within a few tens of AU if the binary orbital eccentricity is mild.  While 
planetary system remnants orbiting white dwarfs are likely to be found outside 5\,AU \citep{nor10,
bur02}, hence leaving a relatively narrow region in which such objects may persist at HD\,27442B, 
their signatures are sometimes found via heavy element pollution in an otherwise pure hydrogen 
or helium atmosphere \citep{zuc10,far10}.

The spectrum of HD\,27442B was examined by eye for photospheric metal lines observed in
DAZ white dwarfs of similar effective temperature; calcium K absorption being by far the most 
prominent for a typical polluted white dwarf \citep{zuc03}.  Despite the excellent data quality,
the search was unsuccessful.  The S/N was estimated to be around 80 in a 300 pixel, 4\,\AA \ 
interval around the calcium K-line (3933.7\AA).  This estimate comes from the raw, unbinned, 
extracted spectrum and uncorrected for the highly curved shape in this region, and hence the 
true S/N is certain to be higher.  Based on comparable but lower S/N UVES observations of 
similar DA white dwarfs with and without metals, a calcium abundance of $\log\, [n({\rm Ca})/
n({\rm H})]>-9.0$ can be firmly ruled out \citep{koe05}.

\subsection{Spectral Energy Distribution}

Figure \ref{fig3} plots the {\em GALEX}, $V$- and $H$-band photometry for HD\,27442B 
\citep{mug07,mar05,mas01}, together with the spectral model fitted to the UVES data.  The 
{\em GALEX} data are uncorrected for extinction, which should be mild at a distance of 18\,pc, 
and the single measurements for each component and each bandpass were assigned 10\% 
errors based on this uncertainty (the quoted errors are less than 1\%, which is unrealistic).  
It is not clear from the {\em GALEX} catalog documentation whether the photometry of either 
component is problematic (e.g.\ too bright), or contaminated by its companion, hence the error 
assignment is also based on this additional uncertainty.  As can be seen in Figure \ref{fig3}, 
the ultraviolet flux of the optically-derived model disagrees somewhat with the {\em GALEX}
photometry, even with the greatly augmented error bars.  Possible reasons for this are beyond 
the scope of the current work.

%%%FIGURE4%%%

\subsection{Evidence for Orbital Motion?}

Astrometric observations of HD\,27442AB exist as early as 1930 and 1964 \citep{mas01}.  
Table \ref{tbl3} lists those data together with relatively recent, near-infrared adaptive optics 
astrometry for the pair.  Although the primary star is badly saturated in the 1\,sec exposure
$V$-band image obtained at CTIO (Figure \ref{fig1}), two methods were used to obtain a 
centroid.  First, the diffraction spikes resulting from the secondary mirror supports were fitted 
with lines whose intersection should coincide with the astrometric center of the star.  Second, 
a weighted centroid was measured for hundreds of linear pixels after removal of non-linear 
and saturated data.  Both of these methods produced agreeable results for the primary star, 
to within a single pixel, and the average offset resulting from those values is listed in Table 
\ref{tbl3}.

%%%TABLE3%%%
\begin{table}
\begin{center}
\caption{Astrometric Observation Summary for HD\,27442AB\label{tbl3}} 
\begin{tabular}{@{}lccr@{}}
\hline
\hline

Epoch		&Separation			&P.A.			&Reference\\
(yr)			&(arcsec)				&(deg)			&\\
 
\hline

1930.1$^a$   	&13.68				&33.9   			&\citealt{jes55}\\
1964.9$^a$   	&13.82				&36.3   			&\citealt{hol66}\\
2003.7$^b$	&$12.94\pm0.08$		&$36.7\pm0.4$		&\citealt{mug07}\\
2004.7$^c$ 	&$13.07\pm0.02$   		&$36.4\pm0.2$  	&\citealt{cha06}\\
2004.8    		&$12.8\pm0.2$ 		&$36.3\pm0.9$		&CTIO $V$-band\\
2007.8   		&$14.0\pm0.2$			&$35.0\pm0.8$		&{\em GALEX} NUV\\

\hline
\end{tabular}
\end{center}

$^a$ The details of the earliest observations lack were provided by B. Mason 
(2010, private communication), and do not have error determinations.
$^b$ Average epoch and offsets for 2 observations\\
$^c$ Average epoch and offsets for 3 observations\\

\end{table}

The {\em GALEX} near-ultraviolet pipeline frame has by far the lowest contrast between
components among any available image of the binary, and computing reliable centroids 
from this image was straightforward.  However, the measured separation in that image (Table 
\ref{tbl3}) disagrees significantly with three other recent measurements, whose mean is $12
\farcs93\pm0\farcs14$ at a mean epoch of 2004.4.  The {\em GALEX} pixel scale is large at 
$1\farcs5$, and while \citet{mor07} report a $1\sigma$ error of $0\farcs5$ in {\em absolute} 
position for bright point sources within the central $0\fdg6$ of the near-ultraviolet image fields 
(which applies to HD\,27442), this should not affect the {\em relative} astrometry for two nearby 
point sources.  Given this notable deviation from the other measures, the {\em GALEX} offsets 
may not be accurate.

The 1930 and 1965 observations give an average projected separation near 250\,AU and
thus $P\ge2665$\,yr for a total binary mass of 2.2\,$M_{\odot}$.  A face-on orbit is unlikely 
given 1) the radial velocity-detected planet at the primary (assuming planet-binary coplanarity
as the most probable configuration), and 2) an expected $\dot \theta=0\fdg090-0\fdg135$\,yr$
^{-1}$ orbital motion for $e=0-0.5$, or $3\fdg6-5\fdg4$ over 40\,yr that has not been observed.  
Thus, it is likely the binary has $i\ga70\degr$.  A circular, edge-on orbit is consistent with a 
$0\farcs8$ change in separation (if real) between 1964 and 2004.  For this case, one gets 
$a=385$\,AU and $P=5090$\,yr by numerically solving orbital equations and Kepler's third 
law simultaneously, suggesting the 2004 projected separation would be around $0.6a$.  
Without precise and accurate astrometry over decade timescales, further constraints on 
the binary orbit are unlikely to be forthcoming.	

While the wide binary orbit is largely unconstrained, these rough estimates indicate highly 
inclined orbits are unlikely (consistent with the radial velocity-detected planet), while orbits 
edge-on yield likely semimajor axes a bit larger than 1.6 times the current projected separation.

\subsection{Planet Formation in the Former A/F Star Binary}

Table \ref{tbl1} lists the resulting main-sequence parameter constraints for HD\,27442B based 
on current mass estimates of the primary star.  In this case, there is poor agreement between
the age estimate of HD\,27442A and the predicted main-sequence lifetime plus cooling age of 
its companion.  Both the initial-to-final mass relation \citep{wil09}, and main-sequence lifetimes
\citep{hur02} are steep functions as stellar mass decreases below 1.6\,$M_{\odot}$, and thus if
the mass of the primary is a few percent lower than the adopted value, the potential for better
agreement between the total ages of the components grows rapidly.  If the adopted white dwarf 
and progenitor masses for HD\,27442B are essentially correct, then its main-sequence lifetime 
was only 1.3\,Gyr and the total system age is under 1.5\,Gyr

With both stars on the main-sequence, the semimajor axis of the binary was shorter by a factor 
of 

\begin{equation}
\frac{a_0}{a} =  \frac{M_{\rm B} + M_A}{M_{\rm ms} + M_A} 
\label{eqn3}
\end{equation}

\begin{figure}
\includegraphics[width=84mm]{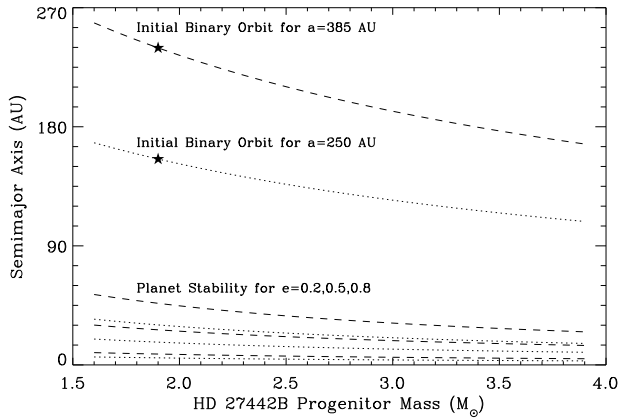}
\caption{Possible semimajor axes for HD\,27442AB during the epoch of planet formation,
plotted as a function of the progenitor mass of the white dwarf secondary.  The star symbol 
denotes the value resulting from the adopted white dwarf mass in Table \ref{tbl1}.  Also shown 
is the critical semimajor axis beyond which planetary orbits are unstable; this depends strongly 
on the binary eccentricity which is unknown but likely to be substantial given $P>10^3$\,yr
(\S3.5).
\label{fig5}}
\end{figure}

\noindent
This ratio is not very sensitive to the spread in possible values for each mass, and gives an 
average value of $a_0=0.62a$ for the various possibilities based on this work and the published 
mass estimates of the primary \citep{tak07,but06}.  If the current projected separation reflects the 
current semimajor axis, then $a_0\approx150$\,AU.  Figure \ref{fig5} plots the resulting orbital 
constraints when the pair were both on the main sequence, both for the stellar and planetary 
companions.  The critical radius for planet stability \citep{hol99} shown in figure corresponds 
to the $a_0$ derived here.

%%%FIGURE5%%%

As a pair of main-sequence stars near 1.5 and 1.9\,$M_{\odot}$, the likely spectral types of 
HD\,27442B and A should have been around A5 and F0, respectively \citep{dri00}.  For their 
initially wide binary orbit, strong constraints on planetary orbit stability arise only for significant 
binary eccentricities.  For the intermediate mass stars in question with an initial period $P>10^3
$\,yr, a mean eccentricity of 0.5 is expected \citep{abt05}, and therefore planetary orbit stability 
would have been confined to within $15-20$\,AU.

\subsection{Total Age}

Using the stellar evolutionary formulae of \citet{hur02}, a main-sequence lifetime of $2.3-
2.8 $\,Gyr results for a star with a mass in the range $1.49-1.59$\,$M_{\odot}$ as estimated 
for HD\,27442A \citep{tak07,but06}.  Given that the primary in this system has recently left the 
main-sequence, and that subsequent evolutionary phases are much shorter lived than the 
hydrogen burning lifetime, the above is a good estimate for the total lifetime of the binary 
and planet-host system.  However, for main-sequence masses near 1.9\,$M_{\odot}$, the 
total age of the white dwarf should be less than 1.5\,Gyr.  If the 2.8\,Gyr age estimate of 
\citet{tak07} is accurate, this would argue for a longer progenitor lifetime for HD\,27442B,
and in the direction of the lower white dwarf mass as derived from spectroscopy.

\section{CONCLUSIONS}

The optical spectrum of the white dwarf companion to the planet hosting star HD\,27442A
yields a reliable effective temperature, and three, mostly independent determinations of its 
radius and mass.  These results combine to support a picture whereby planets orbiting the
HD\,27442A within tens of AU have always been stable, but indicate a total system age that 
is significantly younger than previous estimates.  The white dwarf HD\,27442B does not 
exhibit photospheric metals indicative of a remnant planetary system, but its more massive 
progenitor was capable of hosting stable planets within tens of AU.

Assuming the star-star and star-planet planes are all co-aligned leads to likely binary orbits 
near edge-on, but substantial uncertainty remains.  If orbital motion is apparent in the past 40\,yr, 
the semimajor axis is larger than the current projected separation, leading to current periods up 
to 5100\,yr.  On the other hand, if the projected separation reflects the semimajor axis, the period
is closer to 2500\,yr.  From these estimates, it is found that the binary system was no closer than
150\,AU at the time of planet formation, and only a high eccentricity should have affected stable
planetary orbits within 20\,AU at HD\,27442A.

Future, precision astrometric measurements should be able to confirm or rule out orbital motion
on decade timescales, while ongoing studies of the exoplanet can similarly identify any trend in 
the radial velocity of the primary at the 10\,m\,s$^{-1}$ level.  Such observations would place the
best possible constraints on the binary semimajor axis.  Optical and near-infrared photometry at
the few percent level would better constrain the effective temperature and radius of the white 
dwarf, providing a more precise estimate of the binary separation during the epoch of planet
formation.

\section*{ACKNOWLEDGMENTS}

J. Farihi thanks T. Henry and D. Raghavan for sharing their optical images of HD\,27442.  The 
authors thank the referee G. Chauvin for comments which improved the quality and clarity of the 
manuscript.  D. Koester kindly provided his white dwarf atmospheric models for spectral fitting
(Balmer lines in the models were calculated with the modified Stark broadening profiles of 
\citet{tre09} kindly made available by the authors).  J. Farihi acknowledges the support of the 
STFC.  J. B. Holberg wishes to acknowledge support from NSF grant AST-1008845.  Based on 
observations made with ESO Telescopes at Paranal Observatory under program 382.D-0186(A).  
This work includes data taken with the NASA {\em Galaxy Evolution Explorer}, operated for NASA 
by the California Institute of Technology under NASA contract NAS5-98034.

\label{lastpage}

\end{document}